\documentstyle[12pt]{article}
\newcommand {\bequ}{\begin{equation}}
\newcommand {\eequ}{\end{equation}}
\newcommand {\beq}{\begin{eqnarray}}
\newcommand {\eeq}{\end{eqnarray}}

\def\la{\lambda}

\def\pd{\partial}
\def\a{\alpha}

\def\e{\epsilon}

\begin{document}
\setlength{\oddsidemargin}{0cm}
\setlength{\baselineskip}{7mm}

\begin{titlepage}
 \renewcommand{\thefootnote}{\fnsymbol{footnote}}
$\mbox{ }$
\begin{flushright}
\begin{tabular}{l}
hep-th/0004172\\
KUNS-1661\\
\end{tabular}
\end{flushright}

~~\\
~~\\
~~\\

\vspace*{0cm}
    \begin{Large}
       \vspace{2cm}
       \begin{center}
         {\bf Dirichlet Boundary Conditions in Generalized Liouville Theory
              toward a QCD String}      \\
       \end{center}
    \end{Large}

  \vspace{1cm}

\begin{center}
           {\bf Shin NAKAMURA}$^{1)2)}$\footnote
           {e-mail address : nakamura@gauge.scphys.kyoto-u.ac.jp}

        $^{1)}$ {\it The Graduate University for Advanced Studies,}\\
               {\it Tsukuba 305-0801, Japan} \\
        $^{2)}${\it Department of Physics, Kyoto University,
                   Kyoto 606-8502, Japan}\\

\end{center}

\vfill

\begin{abstract}
\noindent

We consider bosonic noncritical strings as QCD strings and
we present a basic strategy to construct them in the context
of Liouville theory. 
We show that Dirichlet boundary conditions play important
roles in generalized Liouville theory.
Specifically, they are used to stabilize the
classical configuration of strings and also utilized to treat
tachyon condensation in our model.
We point out that Dirichlet boundary conditions for the Liouville mode
maintains Weyl invariance, if an appropriate condition is satisfied, 
in the background with a (non-linear) dilaton.

\end{abstract}

\vfill
\end{titlepage}
\vfil\eject

\section{Introduction}

Gauge theories and string theories are very important fields of modern 
theoretical physics. Construction of a non-perturbative formulation of 
string theories is one of the main subjects which has not yet been
completely understood. Although non-perturbative formulations of gauge
theories were constructed as lattice gauge theories, we are still
interested in analytical studies on non-perturbative effects of gauge
theories.
The relation between gauge theories and string theories has been
studied for a long time, and both theories have developed 
while affecting each other deeply. 
Thus, studies on the connection between them
seems to be important and is expected to be fruitful.
With this in mind, we consider QCD strings in
this article, from the viewpoint of Liouville theory.

Investigation of non-perturbative effects of QCD (for example,
 confinement of quarks)
has been one of the motivations for string theories. In the
context of QCD strings, confinement of quarks is explained in terms of
 a linear 
potential produced by the ``string'' stretched between quark
and antiquark. There are many reasons why we consider such a string 
theory, although this naive explanation has not been completed nor
proved to this time.\footnote{For a review of these topics, see
 Ref. \cite{Polchinski:1992vg}.}

In the strong coupling expansion of lattice gauge
 theories, the quark-antiquark potential is given by the
 expectation value of the Wilson loop, each term of which corresponds
 to different configuration of the lattice surfaces.
This reminds us of the sum of the ``world-sheet surfaces'' of an open
 string whose boundary is the Wilson loop.
We can find another description of gauge theories in terms of strings in
the large-$N$ limit.
In large-$N$ gauge theories, Feynman diagrams are classified by
 the ``Euler number,''
 and the physical observables are expressed as a series with respect 
 to the ``topology'' of the graph.
The dominant terms correspond to planar diagrams. 
This also reminds us of a ``world-sheet'' description.
Furthermore, in the description of the dual Meisner effect, the
 fluxes of color electric charge are collimated, and form a ``string''
 between quarks.
We also find similar correspondence between gauge theories and string 
 theories in the contexts of D-branes and AdS/CFT.\footnote
{We have found other connections between gauge
  theories and strings \cite{Polchinski:1992vg}.}

In consideration of the above discussion, we study QCD strings with
 the following assumptions.

(1) We assume that the color charge is attached to the ends of
 open strings, and we regard the world-sheet boundaries as Wilson loops.

(2) The Wilson loops are not dynamical, since we attempt to describe
pure YM theories in which quarks are not dynamical.

(3) We assume that the strings are bosonic and exist in
four-dimensional spacetime, because our purpose is to describe
four-dimensional (large-$N$) non-SUSY gauge theories.

\section{Review of Liouville theory}

It is most natural to assume that QCD strings
 are bosonic and four-dimensional objects. 
Therefore, we must to find some nontrivial way to construct
 a consistent noncritical string theory.

Quantization of noncritical strings has been considered in Liouville
theories \cite{Polyakov 1981,O.Alvarez}. 
In particular, strings for $d\leq1$ have been consistently quantized
using DDK theory presented by Distler and Kawai \cite{DK} and
David \cite{David}. (Here $d$ denotes the spacetime dimension in
which strings exist.)
For this reason, we first review the work on the DDK theory.\footnote
{For a review of DDK, see Ref. \cite{review of DDK}.}

\subsection{DDK theory without boundaries}

Let us start with a $d$-dimensional ($d\not=26$) bosonic string without
 boundaries for simplicity.
We use the Euclidean signature both for the world-sheet metric and for the 
 spacetime metric.
The Polyakov action is
\beq
S_{M}=\frac{1}{4\pi\alpha '}
\int d^{2}\xi \sqrt{g}(g^{ab}\partial _{a}X^{\mu }\partial _{b}X_{\mu }
),
\label{action}
\eeq
where $\mu$ runs from $1$ to $d$.
The partition function with respect to this action is diffeomorphism
 invariant, and we can fix the world-sheet metric as
\beq
g_{ab}=\hat{g}_{ab} e^{\phi(\xi)}.
\eeq
However, the Weyl invariance of the partition function is broken
 for $d\not=26$ at the quantum level,
 and the freedom of the Weyl transformation
is no longer a gauge freedom. Therefore, we have to perform the path
integral with respect to $\phi$ rigorously.
 However, the measure
$[d\phi]_{g}$ of the path 
integral with respect to $\phi$ is given by the norm
\beq
\|\delta \phi \|_{g}=\int d^{2}\xi \sqrt{g}(\delta\phi)^{2}
=\int d^{2}\xi \sqrt{\hat{g}}e^{\phi(\xi)}(\delta\phi)^{2},
\eeq
which depends on $\phi$ itself in a complicated manner.
Thus, to perform the path integral with this measure is difficult and
 seems to be almost impossible.

In Louville theory based on the DDK argument, the measure of the path
 integral is
redefined with respect to some fixed world-sheet metric $\hat{g}_{ab}$ 
to avoid the above described difficulty. Namely, we
use the measure $[d\phi]_{\hat{g}}$ given by the norm
\beq
\|\delta \phi \|_{\hat{g}}=\int d^{2}\xi \sqrt{\hat{g}}(\delta\phi)^{2},
\eeq
which does not depend on $\phi$.
With this redefinition, we have to use a Jacobian $J$ to maintain
consistency:
\beq
[dX]_{g}[db]_{g}[dc]_{g}[d\phi]_{g}
&=&[dX]_{\hat{g}}[db]_{\hat{g}}[dc]_{\hat{g}}[d\phi]_{\hat{g}} J,
\label{anomaly1}
\eeq
\beq
J&=&e^{-S_{L}},
\eeq
where $b$ and $c$ are the ghost fields.
We have the relations \cite{Polyakov 1981}
\beq
[dX]_{g}[db]_{g}[dc]_{g}
=[dX]_{\hat{g}}[db]_{\hat{g}}[dc]_{\hat{g}} e^{-S_{{\rm Jac}}}
\label{anomaly2}
\eeq
\beq
S_{{\rm Jac}}=\frac{1}{2} \frac{26-d}{48\pi}
\int d^{2}\xi \sqrt{\hat{g}}\{
\hat{g}^{ab}\partial _{a}\phi\partial _{b}\phi+2\hat{R}\phi
\},
\label{S_Jac}
\eeq
where $\hat{R}$ is the world-sheet scalar curvature with respect to
 the metric $\hat{g}_{ab}$. (Note that (\ref{anomaly2}) does not contain
$[d\phi]$, while (\ref{anomaly1}) {\it does} contain $[d\phi]$.)
Therefore, the Jacobian $J$ can be naturally assumed to be
\begin{equation}
S_{L}=u\int d^{2}\xi \sqrt{\hat{g}}\{
\hat{g}^{ab}\partial _{a}\phi\partial _{b}\phi+q\hat{R}\phi
\},
\end{equation}
where $u$ and $q$ are some constants.


Then, the total action, including the Jacobian term, is
\beq
S &=& S_{M}+S_{L}+S_{{\rm ghost}} \nonumber \\
&=& \! \frac{1}{4\pi\alpha'}
\! \int \!\!  d^{2}\xi\sqrt{\hat{g}} \{
\hat{g}^{ab}\partial_{a} \varphi\partial_{b}\varphi
+\hat{g}^{ab}\partial_{a} X^{\mu}\partial_{b}X_{\mu}+\alpha'Q\hat{R}\varphi
\}
+S_{{\rm ghost}},
\label{actionwoc}
\eeq
where we have redefined the factor $q$ and the field $\phi$ as
\beq
Q=\sqrt{\frac{4\pi u}{\alpha '}} \: q \:,
\eeq
\beq
\varphi(\xi)=\sqrt{4\pi\alpha'u}\: \phi(\xi)=\frac{q}{Q\a'}\: \phi(\xi).
\eeq

In the action (\ref{actionwoc}), the field $\varphi$ has a kinetic
 term and can be regarded as a new ``coordinate'' of the target
 space\footnote{Note that $\varphi$ has 
 the dimension of length, though $\phi$ is a dimensionless field.}; 
we obtain one more
spacetime dimension and a linear dilaton term through the
process of quantization.
The world-sheet metric $g_{ab}$ has been replaced with $\hat{g}_{ab}$
in the process.

Let us consider the partition function $Z$ with respect to the action
 (\ref{actionwoc}),
\beq
Z=\int [d\varphi][dX] e^{-S},
\label{pathwob}
\eeq
where $[d\varphi]$ and $[dX]$ stand for
 $[d\varphi]_{\hat{g}}$ and $[dX]_{\hat{g}}$.
(We omit $\hat{g}$ from expressions of the measure in the following.) 
Here $[dX]$ represents
the measure with respect to the matter and the ghost. It also
represents the measure for the modular
integration if $N_{g}>0$, where $N_{g}$ is the number of the genus of the
 world-sheet. 

We note that the partition function does not change
under the simultaneous transformations that keep  
$g_{ab}=\hat{g}_{ab} e^{\frac{Q\a'}{q} \varphi}$
 invariant,
\beq
\hat{g}_{ab}\mapsto \hat{g}_{ab} e^{\delta(\xi)},
\label{Weyl}
\eeq
\beq
\varphi(\xi)\mapsto \varphi(\xi)'=\varphi(\xi)-\frac{Q\a'}{q}\: 
\delta(\xi),
\label{shift}
\eeq
if the boundary condition of the integration in
(\ref{pathwob}) allows the shift of $\varphi$ given by (\ref{shift}).
This is because the action in the initial formulation (\ref{action}) 
depends only on $g_{ab}$ and $X^\mu$, which do not change under the above
transformations. Thus,
\beq
Z=\int [d\varphi][dX] e^{-S[\hat{g},X,\varphi]}
=\int [d\varphi '][dX] e^{-S[\hat{g} e^{\delta(\xi)},X,\varphi ']}.
\eeq
After rewriting the dummy variable $\varphi '$ as $\varphi$,
we note the very important fact that the partition
function is now represented in a Weyl {\em invariant} way
{\em with respect to the new metric $\hat{g}_{ab}$}. 

Now we are in a position to determine the value of $Q$.
We found that the partition function is Weyl invariant. Therefore, the total
central charge of the theory should be zero:
\beq
C_{\varphi}+C_{M}+C_{{\rm ghost}}=0 \: ,
\label{ccancel}
\eeq
where $C_{\varphi}$ is the central charge for $S_{L}$, $C_{M}$ is the
central charge for $S_{M}$, which is equal to $d$, and $C_{{\rm ghost}}$ is
the central charge for $S_{{\rm ghost}}$, which is calculated to be $-26$.
We can evaluate $C_{\varphi}$ with a standard technique of conformal field
theory as
\beq
C_{\varphi}=1+6\alpha 'Q^{2}.
\label{cphi}
\eeq
Thus (\ref{ccancel}) and (\ref{cphi}) give us
\beq
Q=\pm\sqrt{\frac{25-d}{6\alpha'}}\:.
\label{Q}
\eeq

We point out that
 our action is exactly the same as that of the ``linear
dilaton string'' in $d+1$ dimensional spacetime. In the linear dilaton
string, $\varphi$ is regarded as one of the spacetime coordinates, and
 $Q$ is given as a factor which appears in
the dilaton term. However, from the viewpoint of Liouville theory,
 we do not regard $\varphi$ as a real physical
  coordinate of spacetime. It is the parameter of the Weyl 
 transformation and becomes the new spacetime
 coordinate through the quantization process. Thus, we call
 $\varphi$ (or $\phi$) the ``Liouville mode'' in this article.

Now we have some problems.
First, the Liouville mode $\varphi$ does not have a stable vacuum 
with the action (\ref{actionwoc}). 
In Refs. \cite{DK} and \cite{David}, 
a cosmological
constant is added to the action to obtain a
 stable vacuum. The Polyakov action with the (renormalized) cosmological
constant $\mu$ is
\beq
S_{M}=\frac{1}{4\pi\alpha'}
\int d^{2}\xi \sqrt{g}(g^{ab}\partial _{a}X^{\mu }\partial
_{b}X_{\mu }+\alpha'\mu).
\eeq
It is natural to
assume that the cosmological constant term is deformed by the quantum effect
as
\beq
\frac{\mu}{4\pi}\int d^{2}\xi \sqrt{\hat{g}} e^{\phi} 
&\longmapsto&
\frac{\mu}{4\pi}\int d^{2}\xi \sqrt{\hat{g}} e^{\gamma\phi} \nonumber \\
& &=
\frac{\mu}{4\pi}\int d^{2}\xi \sqrt{\hat{g}} e^{\alpha \varphi},
\eeq
where $\gamma$ is a constant indicating an anomalous dimension, and we
defined $\alpha$ as $\alpha\equiv \frac{Q\a'}{q} \gamma $.
To preserve the Weyl invariance of the theory, we choose
$\alpha$ so that the cosmological constant term is Weyl
invariant. This is realized if the conformal dimension of the 
operator $e^{\alpha\varphi}$ is 2. Namely, we demand $e^{\alpha\varphi}$
to be a (1,1)-primary operator.
In complex coordinates, the holomorphic part of the energy-momentum
 tensor given by the
 action (\ref{actionwoc}) is
\beq
T_{ZZ}=-\frac{1}{\alpha'} (\partial\varphi\partial\varphi
-Q\partial^{2}\varphi),
\eeq
and the conformal weight $\Delta$ of $e^{\alpha\varphi}$ is given by 
\beq
\Delta=\frac{\alpha'}{2} \alpha\left (Q-\frac{\alpha}{2}\right),
\label{Delta}
\eeq
which should be 1. Thus we obtain two solutions:
\beq
\alpha_{\pm}=Q \pm \sqrt{Q^{2}-\frac{4}{\alpha'}}
=\sqrt{\frac{25-d}{6\alpha'}}\pm \sqrt{\frac{1-d}{6\alpha'}}\:.
\eeq
In the classical limit ($d\rightarrow -\infty$), $\alpha$ should be zero, 
and we take the branch of $\alpha_{-}$. Then we have a stable
vacuum for a negative world-sheet curvature $\hat{R}$.
This is because the potential for $\varphi$ is given by
\beq
V(\varphi)=\frac{1}{4\pi}
\int d^{2}\xi (Q \hat{R} \varphi + \mu e^{\alpha\varphi}),
\eeq
and has a minimum value if the factor of $\varphi$ in the first term
is negative.
(We have assumed a constant-$\varphi$
configuration as a classical solution here.)

We have a second problem with the strings for $d>1$. In this region,
 $\alpha$ is a complex number. Thus we have to regard
$\mu e^{\alpha\varphi}$
as a tachyon vertex operator with momentum in the $\varphi$ direction,
rather than 
a cosmological constant term.
Furthermore, the composite operator
$e^{\alpha\varphi}$ becomes non-normalizable \cite{Seiberg}.
We also have to consider the condensation of the 
target-space tachyons, since we treat a noncritical bosonic string in which the
tachyonic mode is not projected out. 
We point out that we do not have target-space tachyons if $d\leq1$.
This is because the Liouville theory is described as a $d+1$-dimensional 
 string theory, and the world-sheet oscillation
 can be fixed completely by the gauge symmetry if $d+1 \leq 2$.
Thus the tachyonic mode cannot appear.

For this reason, we can construct a Weyl invariant string theory with
a cosmological constant for the case $d\leq1$. A consistent model
for quantized noncritical strings for $d>1$ has not yet been constructed.

\subsection{Liouville theory with boundaries}
For noncritical strings with a boundary, we assume $S_{L}$ to be
\bequ
S_{L}
=\frac{1}{4\pi\alpha'}
 \int_{{\cal M}} d^{2}\xi \sqrt{\hat{g}}\{
\hat{g}^{ab}\partial _{a}\varphi\partial _{b}\varphi
+\alpha'Q\hat{R}\varphi
\}
+\frac{1}{2\pi\alpha'}
\int_{\partial {\cal M}} ds\sqrt{\hat{g}_{ss}}
\alpha'Q\hat{k}\:\varphi \:,
\eequ
where $s$ is a parameter of the boundary $\partial {\cal M}$ and
 $ds \sqrt{\hat{g}_{ss}}$
 denotes an invariant infinitesimal length on it \cite{O.Alvarez}.
The quantity $\hat{k}$ is the
extrinsic curvature with respect to the metric $\hat{g}_{ab}$.
If the world-sheet has many boundaries, $\partial {\cal M}$ denotes
all of them.
 The only difference between this and the boundaryless case is the 
existence of the boundary terms. 
We can also add a boundary cosmological constant term, 
\beq
\mu_{b}\int_{\partial {\cal M}} ds\sqrt{\hat{g}_{ss}} e^{\frac{\alpha}{2}
  \varphi}.
\eeq

For strings with boundaries, we also need to consider the boundary
 conditions.
We can consider several types of boundary
conditions for $\varphi$.\footnote
{Boundary conditions and boundary states for linear dilaton theory
 have been considered by several authors. (For example, see
 Refs. \cite{O.Alvarez},\cite{boundary state},\cite{Polchinski 1994}
and references therein.)}
Now we set $\hat{k}=0$ for simplicity. The reason why this choice is
natural is discussed in \S 3.1.
In this case, we have two choices for the boundary
conditions. One of them is Neumann boundary conditions, and the other
is Dirichlet boundary conditions.

Neumann boundary conditions allow the ends of
open strings to move freely. In other words, Neumann boundary 
conditions can be regarded as an equation of motion for the endpoint.
 Therefore, it does not restrict the discussion in \S 2.1, and
 the argument given for the boundaryless case still holds.
 On the other hand, we have to be careful if we consider
Dirichlet boundary conditions. This is because the shift of
 $\varphi$ (\ref{shift}) is not consistent with Dirichlet boundary
 conditions, and thus Weyl invariance is broken, \cite{Polchinski 1994}
 in general.

However, we show in \S 3.3 that we can use
Dirichlet boundary conditions without breaking Weyl invariance
in some special case.

\section{Liouville theory as QCD string --- \\
--- A trial to go beyond the {\boldmath $d=1$} barrier}
\setcounter{equation}{0}

\subsection{Liouville theory as a QCD string}

We presented our basic assumptions for noncritical QCD strings in
\S 1. Here we make these more precise before proceeding with a further
 argument.

The boundaries of the world-sheet correspond to non-dynamical
 rigid Wilson loops.
In this sense, the boundary conditions for $X^{\mu}$ should be Dirichlet 
boundary conditions.
A nontrivial problem is how to choose the boundary conditions for $\varphi$.

We choose the topology of the classical world-sheet
 to be a cylinder, and we set $\hat{k}=0$ for simplicity.
This is reasonable for the calculation of the static quark-antiquark
 potential. We usually consider a rectangular Wilson
 loop of infinite length along the time direction to calculate it.
 However, $\hat{k}$
 diverges at the corners of the rectangle, and this makes calculations
 difficult.
 To avoid this, we connect the shorter sides (the space-like sides) of
 the rectangle and thereby make it periodic, like a ring. This configuration
 consists of two parallel circular Wilson loops, and is like a
 cylinder.
 Then, the corners disappear, and the
 boundaries are straight from the two-dimensional viewpoint on the
 world-sheet. Then we can set
 $\hat{k}=0$ naturally on the world-sheet.
 Of course, the color
 charges on the loops are set to make a color singlet. In the last
 step of the calculation, the periodicity of the
 loops is set to infinity, and thus the calculated value of the
 potential should attain the same value as that for an
 infinitely long rectangular Wilson loop.

Next, we make our problems clear.
To go beyond the $d=1$ barrier, we have mainly two problems to
solve. 
First, we must stabilize the vacuum of the field
$\varphi$, or fix the zero mode of $\varphi$ to
stabilize the classical configuration of the string.
Second, we have to treat the condensation of the 
 target space tachyons if they exist.

We propose some basic ideas to solve these problems in the 
following sections.
We find that Dirichlet boundary conditions play important roles
in the generalized Liouville theory.

\subsection{Generalization of Liouville theory}

We saw that the cosmological constant term becomes tachyonic for
 $d>1$. Thus we cannot use it to stabilize $\varphi$.

One idea to stabilize $\varphi$ without a cosmological constant is
to add another Weyl invariant term to the action, which generates
the minimum of $V(\varphi)$. We point out that one of the simplest
 candidates for such a term is
\beq
\frac{\mu'}{4\pi\alpha'}\int d^{2}\xi \, e^{2Q\varphi}
\hat{g}^{ab}\partial _{a}X^{\mu }\partial _{b}X_{\mu },
\label{ads}
\eeq
where $\mu'$ in (\ref{ads}) is a constant. 
This is because one can verify that the operator
\beq
 e^{2Q\varphi}\hat{g}^{ab}\partial _{a}X^{\mu }\partial _{b}X_{\mu }
\label{adsop}
\eeq
has conformal dimension 2, namely $\Delta(e^{2Q\varphi})=0$ from
 (\ref{Delta}).
Then we obtain a new action with (\ref{ads}):
\beq
S\!= \! \frac{1}{4\pi\alpha '}\!\!
\int \!\! d^{2}\xi\sqrt{\hat{g}} 
\{
(1+\mu' e^{2Q\varphi})
\hat{g}^{ab}\partial_{a} X^{\mu}\partial_{b}X_{\mu}
+\hat{g}^{ab}\partial_{a} \varphi\partial_{b}\varphi
+\alpha 'Q\hat{R}\varphi
\} 
\!+\!S_{{\rm ghost}}. \makebox[1.2cm]{}
\label{actionads}
\eeq
However, one finds that the above action is not exactly 
Weyl invariant.\footnote{Although the composite operator
  (\ref{adsop}) is a (1,1)primary operator, multiple insertion of
  them creates another divergence, which causes breakdown of the scale
  invariance. 
  Thus the insertion of the operator (\ref{adsop}) into the action
  breaks Weyl invariance.

  In the case of a cosmological constant term,
  the multiple insertions do not give any divergence. }

Let us consider the most general action for Liouville theory.
 We naturally impose $SO(d)$ symmetry on $d$-dimensional 
spacetime, and we then obtain
\beq
S =\frac{1}{4\pi\alpha'}
 \int d^2 \xi \sqrt{\hat{g}}\{ a^2 (\varphi) (\pd X^{\mu})^2 +(\pd\varphi)^2+ 
\alpha'R^{(2)}\Phi(\varphi)
\},
\label{actionAG}
\eeq
where we have fixed the $\varphi \varphi$-component of the target-space metric
to be 1, with appropriate redefinition (or target space
general coordinate transformation) of $\varphi$. Now the cosmological
constant term is assumed to be zero. (We have suppressed the ghost term
here. Also, note that we have dropped the boundary terms, since we 
set $\hat{k}=0$.)

The $\beta$-functions which should vanish for Weyl invariance are

\beq
\beta_{MN}^{G}=\alpha'({\cal
R}_{MN}  +2\nabla_{M}\nabla_{N}\Phi)+O(\alpha'^{2}),
\eeq
\beq
\beta^{\Phi}=\alpha' \left ( 
-Q^{2}-\frac{1}{2}\nabla^2 \Phi+(\nabla
\Phi)^2
\right ) + O(\alpha'^{2}),
\eeq
where $M$ and $N$ run from $1$ to $d+1$, and $X^{d+1}=\varphi$.
 $ {\cal R}_{MN} $ is the Ricci tensor of the
$d+1$-dimensional target space.

Collecting the above results, the equation of motion up to 
order $\alpha'^{2}$ become
\beq
0={\cal R}_{MN} +2\nabla_{M}\nabla_{N}\Phi,
\label{eqR}
\eeq
\beq
0=-Q^{2}-\frac{1}{2}\nabla^2 \Phi  + (\nabla \Phi)^2,
\label{eqPhi}
\eeq
where $Q$ is given by (\ref{Q}).
We stress that we can use these equations only in the region 
$ \alpha'{\cal R} \ll 1$, where the perturbative approximation for
 the $\beta$-functions is valid.

The solution of these equations is given in Ref. \cite{AG} 
as
\beq
a(\varphi)= a_0\sqrt{\frac{1+\la e^{2Q\varphi}}{1 - 
\la e^{2Q\varphi}}} \:,
\label{asol}
\eeq
where $\la$ and $a_0$ are constants, and
\beq
\Phi= Q\varphi- \frac{3}{2}\log( 1 - \la e ^{2Q \varphi})+ 
\frac{1}{2}\log(1 + \la e ^{2Q \varphi}) +\mbox{constant} \: .
\eeq
The allowed region for $\varphi$ is given by $|\la e^{2Q\varphi}|<1$, or
\beq
\varphi<-\frac{1}{2Q}\log|\la|.
\eeq
We have to check the validity of this solution before further
calculations.
 The target space scalar curvature for the solution (\ref{asol}) is
 given by
\beq
{\cal R} &=& -3 \left ( \frac{a'}{a} \right ) ^{2}
-5\frac{a''}{a} \nonumber \\
&=& \frac{-4Q^{2}b}{(1-b^{2})^{2}} (5b^{2}+8b+5)
\longrightarrow
\left\{
\begin{array}{ll}
-\infty & (\la>0) \\
0 & (\la=0) \\
+\infty & (\la<0)
\end{array}
\right.
\: \mbox{as } |b|\rightarrow 1 \:,
\eeq
where $b=\la e ^{2Q \varphi}$ and $a'$ denotes
 $\frac{\partial a}{\partial \varphi}$.
 Therefore, the solution is valid only in the
region satisfying $|b|\ll 1$, or equivalently

\beq
\varphi \ll -\frac{1}{2Q}\log|\la| \:,
\label{regionb}
\eeq
if $\la \neq 0$.
In the case $\la = 0$, the solution is exactly the same as
 the ``linear dilaton string''
considered for $d\leq 1$. We investigate the $\la \neq 0$
solution here.
In the region satisfying (\ref{regionb}), the solution is expanded as
\beq
a^{2}(\varphi)=a_0^{2} \{1+2\la e ^{2Q \varphi}\}+O(b^{2}),
\label{sola}
\eeq
\beq
\Phi(\varphi)= Q\varphi + 2\la e ^{2Q \varphi}+O(b^{2}) .
\label{solPhi}
\eeq
We find that the above solution is consistent with the proposed action
given in (\ref{actionads}), if we set $\mu'=2\la$ and rescale $X$ in
 (\ref{actionAG}) to $\frac{X}{a_{0}}$.

Although the best method is to find a solution for the exact equation
of motion, this seems to be very difficult. Thus we 
consider the physics only in the region defined by (\ref{regionb}), 
where the perturbation with respect to 
$\alpha'$ is valid, and we use the action 
\beq
S&=&\frac{1}{4\pi\alpha '}
\int d^{2}\xi\sqrt{\hat{g}} 
\{
(1+2\la  e^{2Q\varphi})
\hat{g}^{ab}\partial_{a} X^{\mu}\partial_{b}X_{\mu}
+\hat{g}^{ab}\partial_{a} \varphi\partial_{b}\varphi
+\alpha '\hat{R}(Q\varphi+2\la e ^{2Q \varphi})
\} \nonumber \\
& &+ S_{{\rm ghost}}
\label{actionfinal}
\eeq
in this region.

\subsection{Stabilization of the vacuum of {\boldmath $\varphi$}}

In this subsection, we attempt to stabilize the vacuum for $\varphi$
in the action (\ref{actionfinal}).
We have two strategies. One is the standard stabilization using
the minimum of the potential. However, we find that
the potential in which we are interested has no minimum in some cases.
 The other one is to impose
Dirichlet boundary conditions for $\varphi$ and fix its zero mode.
To use this method, we have to check the consistency of the 
Dirichlet boundary conditions and Weyl invariance.

\subsubsection{Stabilization using the potential minimum}
In the case that Dirichlet boundary conditions are not imposed, the 
zero mode of
$\varphi$ must be stabilized at the minimum point of the
potential.

The vacua of $X^{\mu}$ are stable, and only the stabilization of
$\varphi$ is needed in our model. Thus, let us consider the effective
 action which we obtain after the 
 path integration with respect to $X^{\mu}$,
\bequ
S(\varphi)=\frac{1}{4\pi\alpha '}\!
\int \! \! d^{2}\xi\sqrt{\hat{g}} 
\{
(1+2\la  e^{2Q\varphi}) \langle
\hat{g}^{ab}\partial_{a} X^{\mu}\partial_{b}X_{\mu} \rangle
+\hat{g}^{ab}\partial_{a} \varphi\partial_{b}\varphi
+\alpha '\hat{R}(Q\varphi+2\la e ^{2Q \varphi})
\},
\eequ
where $\langle \cal{O} \rangle$ denotes the expectation value of the operator
$\cal{O}$ obtained by the path integration with respect to $X^{\mu}$.
The equation of motion with respect to the zero mode of $\varphi$
is given by
\beq
\delta S
=\delta \varphi_{c} \int d^{2}\xi\sqrt{\hat{g}} v'(\varphi_{c})
=0,
\eeq
where $\varphi_{c}$ is the zero mode of $\varphi$, and $v(\varphi_{c})$ is
 expressed by
\beq
v(\varphi_{c})=\alpha '\hat{R}Q\varphi + 2\la  e^{2Q\varphi_{c}}(\alpha
'\hat{R}
+ \langle \hat{g}^{ab}\partial_{a}X^{\mu}\partial_{b}X_{\mu} \rangle ).
\label{vphic}
\eeq
The Gauss-Bonnet theorem states that
\beq
\int_{{\cal M}} d^{2}\xi\sqrt{\hat{g}} \hat{R}=4\pi\chi \:,
\eeq
where $\chi$ is the Euler number of the world-sheet ${\cal M}$.
It is given by
$\chi=2-2N_{g}-N_{b}$,
 where $N_{g}$ is the number of the genus of ${\cal M}$,
 and $N_{b}$ is the number of
 boundaries of ${\cal M}$.
Then the equation of motion for $\varphi_{c}$ is 
\beq
 4\pi\alpha' Q \chi+4\la Q \e^{2Q\varphi_{c}}( 4\pi\alpha'\chi +
A)=0,
\label{eomvarphi}
\eeq
where
\beq
A=\int_{{\cal M}} d^{2}\xi\sqrt{\hat{g}}
\langle \hat{g}^{ab}\partial_{a}X^{\mu}\partial_{b}X_{\mu} \rangle
\quad >0.
\eeq
Since $\la \neq 0$ and $Q \neq 0$, we obtain
\beq
e^{2Q\varphi_{c}}=-\frac{ 4\pi\alpha'\chi}
{4\la ( 4\pi \alpha'\chi+A )} \quad >0,
\eeq
and we immediately note that we have no solution for $\chi=0$.
We also obtain
\beq
v''(\varphi_{c})=- 8\pi\alpha' Q^{2}\chi
\eeq
from (\ref{vphic}) and (\ref{eomvarphi}), and we have no
stable vacuum for $\chi > 0$.

Note that the leading string diagram in our model is a cylinder,
for which we have $\chi=0$.
In general, the leading term can be a disk
($\chi=1$) or a cylinder in the calculation of the correlation
functions of Wilson loops.
Thus, unfortunately, we do not have a stable vacuum suitable for
 these $\chi \geq 0$ cases.\footnote{We have the same
 problem for $\chi \geq 0$ in DDK. However, in the calculation of string
 susceptibility, insertion of a
 $\delta$-function into the path integral, which keeps the world-sheet
 area constant, allows us to obtain the correct value for $d \le 1$.}

\subsubsection{The Weyl invariant Dirichlet boundary conditions
and stabilization of the vacuum of {\boldmath $\varphi$}}
We have seen in our model that we cannot make a consistent
 potential for $\varphi$ with disk and cylinder
diagrams.
However, if we impose Dirichlet boundary conditions and fix the zero mode 
$\varphi_{c}$, we are free of this difficulty.

The problem in this case is the compatibility of the Dirichlet boundary
 conditions and the Weyl invariance of the world-sheet. In a linear dilaton 
 string, Dirichlet boundary conditions break Weyl invariance
 \cite{Polchinski 1994}.
Therefore, we have to find a method to introduce Dirichlet boundary
 conditions into our theory without breaking Weyl invariance.

A good place to start is to recall the origin of the Weyl-invariance breaking
in a general dilatonic string with Dirichlet boundary conditions. 
In dilatonic string theory, we usually need a freedom of field
 redefinition to preserve Weyl invariance.
For example, we have to make a constant
shift (a field redefinition) of $\varphi$, are in (\ref{shift}), to
cancel the variation caused by 
 the Weyl transformation (\ref{Weyl}) for $d \leq 1$.
 Dirichlet boundary conditions for $\varphi$ forbids such
 a shift at the boundary, and breaks Weyl
 invariance. On the other hand, Dirichlet boundary conditions for
 $X^{\mu}$ do not break Weyl invariance, because no shift of
 $X^{\mu}$ is needed.

However,
{\bf we point out that we can employ Dirichlet boundary conditions
for {\boldmath $\varphi$} in a general dilatonic string if the criterion 
stated below is satisfied}.
In the theory with a dilaton, the required shift of the field is not a
constant in general. For example, the shifts we need for the fields in
(\ref{actionAG}) at the one-loop level are given as
\beq
\delta \varphi=-\frac{\alpha'}{2}\partial_{\varphi}
\Phi(\varphi) \delta\sigma ,
\label{delvarphi}
\eeq
\beq
\delta X^{\mu}=0,
\eeq
where
$\partial_{\varphi}$ stands for 
$\frac{\partial}{\partial \varphi}$.\footnote
{The required field redefinition to preserve Weyl invariance appears
  in articles which discuss the $\beta$-functions for
  non-linear sigma models. For the models with boundaries, see
  Refs. \cite{NLS-open} and \cite{H.Osborn}. 
  The condition (\ref{delvarphi}) for the shift is also found
 in Ref. \cite{Polchinski 1994}.
}
We can check that (\ref{delvarphi}) gives the 
correct constant shift required in the linear dilaton case
(\ref{shift}) if we set $q=2$.\footnote
{If $q=2$, we get $u=\frac{1}{2}\frac{d-25}{48\pi}$. The
  denominator naturally coincides with the denominator of the factor
  in (\ref{S_Jac}).
 In this linear dilaton case, the Weyl anomaly is
  obtained exactly at the one-loop level.} 
We look deeper into the origin of the field redefinition in the Appendix.

We note that we do {\em not} need field redefinition for $\varphi$ if
$\partial_{\varphi}\Phi=0$.
Therefore, we can use the Dirichlet boundary conditions
\beq
\varphi|_{\partial {\cal M}}=\varphi_{0}\:,
\label{dirichlet_varphi}
\eeq
where
\beq
\partial_{\varphi}\Phi(\varphi)|_{\varphi=\varphi_{0}}=0,
\label{dirichlet}
\eeq
{\em without breaking Weyl invariance} at the one-loop level.
The above stated criterion for consistent Dirichlet boundary
conditions is one of
 the most important assertions of this article.

As a next step, let us examine whether our model (\ref{actionfinal})
has such a $\varphi_{0}$ or not.
Our dilaton term is
\beq
\Phi(\varphi)=Q\varphi+2\la e ^{2Q \varphi}+O(b^{2}).
\eeq
Thus the condition (\ref{dirichlet}) is
\beq
\partial_{\varphi}\Phi=Q+4\la Q e^{2Q \varphi}+O(b^{2})=0,
\label{ourdirichlet}
\eeq
and we obtain
\beq
\varphi_{0}=-\frac{1}{2Q} \log (-4\la)+O(b^{2})
\eeq
for $\la<0$. 
The condition (\ref{ourdirichlet}) is valid only in the region
satisfying $|b|=|\la e^{2Q \varphi}| \ll 1$, and now
 $\la e^{2Q \varphi_{0}}=-1/4$. Thus the above result suggests that we
 have an appropriate point $\varphi_{0}$ for the Dirichlet boundary
conditions (\ref{dirichlet_varphi})
if $\la<0$.

For this reason, we choose $\la<0$ and impose Dirichlet boundary
conditions on the ends of the string to stabilize its
configuration for arbitrary $\chi$.
Note that the above argument naturally selects the branch 
of $\la$ uniquely.

\subsection{Tachyon condensation: another role of Dirichlet
  boundary conditions in Liouville theory}

We found that we can use Dirichlet boundary condition
to stabilize the string configuration, while preserving Weyl invariance
 for arbitrary $\chi$, at least up to $O(\a'^{2})$.

However, we have another big problem:
If the target-space dimension ($d+1$ for our model) is greater than 2,
 we cannot fix all of the freedom of the world-sheet
 oscillation with gauge symmetry, and we have a physical oscillation.
It is well known that we have a tachyonic state in the oscillation
mode of the bosonic string in flat spacetime.
In our model (\ref{actionfinal}),
the target space becomes asymptotically flat in the
 region $\varphi \ll0$.
Now we are considering the case $d>1$.
Thus a tachyonic ground state appears in the region in which the
spacetime is almost flat.
If we have a tachyonic mode, we have to handle tachyon condensation.

Tachyon condensation has been discussed by many authors (for example,
see Ref. \cite{Sen}
), but this 
subject is difficult and has not yet been solved completely.
We present some ideas to treat tachyon condensation here \cite{Kawai}.
We stress that Dirichlet boundary conditions play an important role in 
this subsection too.

First, let us recall field condensation in usual quantum
field theories. In field theory with field condensation (that is, 
in the theory with nonzero expectation value of the field ), we can
calculate correct quantities if we know the correct expectation value 
of the field, even with the perturbation around an incorrect vacuum.
For example, we can calculate the exact propagator with the perturbation
 around an incorrect vacuum by attaching tadpole diagrams to the tree
 propagator. Even though the mass
 squared is negative in a description around such a
 vacuum, tadpoles with appropriate weight create an additional shift
 of the mass squared, and make the total mass squared positive.
In such a case, although the tachyonic mode exists in a perturbative 
theory around an incorrect vacuum, the theory is never wrong, and only
the ``vacuum'' is wrong. In
 field theories, the true vacuum or exact expectation
value of the field can be given by the Schwinger-Dyson equation. Thus we can
get the correct weight of the tadpoles, we can calculate the
correct propagator, and so on.

We now make an analogy between field theories and tachyonic 
string theories. 
Although we have a tachyonic mode, we believe that the string
theory is not fatally flawed, and the problem is that we do not know the true
vacuum of it.
The analogy with field theories tells us that we may be able to obtain
 correct quantities if we attach correct ``tadpoles'' to the
 world-sheet.
 Then, the question is what is the ``tadpole'' in string
 theories.

We guess here that the tadpole in string theories is a macroscopic hole
 with Dirichlet boundary conditions in the world-sheet.
We assume that the Dirichlet boundary condition for them is
\beq
X^{\mu}(\xi_{i})=a_{i}^{\mu},
\label{tad-1}
\eeq
\beq
\varphi(\xi_{i})=\varphi_{0} \: ,
\label{tad-2}
\eeq
where $i$ distinguishes each tadpole, $a_{i}^{\mu}$ is a constant, and
 $\varphi_{0}$ is a constant which satisfies the condition
 (\ref{dirichlet}).
 Of course, we have to consider the proper weights of the string wave
 functions on it.
The above assumption results from the following considerations.

The tachyonic tadpole is an off-shell state, because
it does not carry momentum. We also know that off-shell
states in string theory do not
correspond to local emission vertexes. Thus we naturally
assume that the tachyonic tadpole is a non-local macroscopic hole on
the world-sheet.\footnote
{Some argument for the macroscopic hole as a tachyonic state
 is given in Ref. \cite{Seiberg}.}
We also have to preserve Weyl invariance, and it is
natural to impose the above Dirichlet boundary conditions 
(\ref{tad-1}) and (\ref{tad-2}) on the edge of the hole.
Neumann boundary conditions cannot be taken for a tadpole for
the following reason.
If we impose Neumann boundary conditions at a hole, the value of $X^{\mu}$
changes along the edge of the hole. This means that
we observe a macroscopic hole even in $d$-dimensional spacetime.
Although the Neumann boundary conditions for $\varphi$ do not make a 
macroscopic hole in the visible $d$-dimensional spacetime,
 the hole can have
 momentum in the $\varphi$ direction. This allows us to make an
 on-shell state, and the hole can break into
on-shell open strings moving along the $\varphi$ direction.
These situations do not seem to be natural for our model. 
Contrastingly, the tadpole with the Dirichlet boundary conditions
(\ref{tad-1}) and (\ref{tad-2})
does not leak any momentum from the world-sheet,
 and this gives a natural property for the tadpole.

The macroscopic holes with Dirichlet boundary conditions on the 
world-sheet (or D-instantons in the target space) and the
non-perturbative effects induced by them are discussed in Refs.
\cite{Green et al} and \cite{Polchinski 1994}.
However, we insist that the macroscopic holes discussed here play the
role of ``tadpoles'' naturally even in Liouville theory.

Unfortunately, we do not have the Schwinger-Dyson equation of 
string theory, 
and we do not know how to obtain the correct weight which should be
attached to the tadpole.
Thus, we cannot give a rigorous discussion to treat tachyon
condensation, but we present a
rough argument regarding tachyon condensation.

To treat a macroscopic hole on the world-sheet is rather difficult,
and we therefore approximate it as a point on the world-sheet which 
couples to the Dirichlet boundary conditions.
In this case, the insertion of the tadpole is regarded as the
insertion of
\beq
h\int_{{\cal M}} d^{2}\xi_{1} \sqrt{\hat{g}}\:
\delta(X^{M}(\xi_{1})-a^{M}(\xi_{1}))
\label{tadpole}
\eeq
into the world-sheet, where $h$ is the weight of the tadpole, 
and $\xi_{1}$ denotes the insertion {\em point} on the world-sheet.
We have $X^{d+1}=\varphi$ and $a^{d+1}=\varphi_{0}$.
In usual strings without a dilaton,
the $\delta$-function in (\ref{tadpole}) becomes 1
after the integration over the moduli $a^{M}$, and the string
propagator with the tadpoles can be estimated as
\beq
\mbox{propagator}&=& \frac{1}{L_{0}+\bar{L_{0}}}
+\frac{1}{L_{0}+\bar{L_{0}}} h \frac{1}{L_{0}+\bar{L_{0}}}
+\frac{1}{L_{0}+\bar{L_{0}}} h \frac{1}{L_{0}+\bar{L_{0}}}
h \frac{1}{L_{0}+\bar{L_{0}}}
+\cdots \nonumber \\
&=& \frac{1}{L_{0}+\bar{L_{0}}-h}\:,
\eeq
where $L_{0}$ ($\bar{L}_{0}$) is the (anti)holomorphic part of the
 Hamiltonian of the corresponding conformal field theory.
Thus, the insertion of the tadpoles (\ref{tadpole}) seems to make an
 additional shift to the energy of the tachyonic state.

However, we cannot apply the above estimation directly to Liouville theory. 
Although we should integrate over $a^\mu$ to get Poincar\'e invariance 
in the $d$-dimensional spacetime, we never integrate over
$\varphi_{0}$ in our model (because it is fixed).
Therefore, the expected non-perturbative effects induced by the tadpoles in
Liouville theory seem to be different from those of non-dilatonic 
strings.
In any case, we must develop a technique to estimate 
the effects of the insertion of Dirichlet boundaries.

\section{Conclusion}

We attempted to quantize a noncritical (four dimensional) bosonic string
as a natural candidate of a (large-$N$) pure QCD string.
We considered the generalized Liouville action (\ref{actionAG}) as such a
string.

One of the main problems here is the stabilization of the Liouville mode
$\varphi$ while preserving Weyl invariance, and we found that we can
stabilize it with the Dirichlet boundary conditions (\ref{dirichlet_varphi}).
The criterion for consistent Dirichlet boundary
 conditions at the one-loop level is given by (\ref{dirichlet}), and the
 stabilized point is independent of the topology of the world-sheet.
We found that the perturbative solution 
 for the background in (\ref{actionAG}) seems to have a point which 
 satisfies the condition (\ref{dirichlet}).
This argument led us to the unique selection of the branch of the solution.
 
We also discussed tachyon
 condensation. Although the complete treatment of it is very
 difficult, we
 presented a simple strategy for it. The idea we presented is to
 attach tadpoles to the world-sheet, and we guessed that the tadpole
 in string theories might be represented as a macroscopic hole
 with Dirichlet boundary conditions.
In non-dilatonic cases, a naive approximation for the propagator with
 tadpoles implies a shift of the energy of the tachyonic state.
However, we guess that the non-perturbative effects in
 Liouville theory are different from those of non-dilatonic 
 strings.

Before we close this article, we stress again that Dirichlet boundary
 conditions have an important
role in the generalized Liouville theory, and they can be imposed on the
Liouville mode while preserving Weyl invariance if the appropriate
condition mentioned above is satisfied. The investigation of Dirichlet
strings in dilatonic backgrounds is very important, and it should
yield necessary information about the construction of noncritical
strings.

\begin{center} \begin{large}
Acknowledgments
\end{large} \end{center}

The author would like to thank Professor H.Kawai for valuable and inspiring
 discussion about the subjects presented here.

\appendix
\section{The Origin of the Field Redefinition
 and the Derivation of (\ref{delvarphi})} 
\setcounter{equation}{0}

Here we look deeper into the origin of the necessity of the field
redefinition, at the one-loop level [namely, up to $O(\a'^{2})$ ].
Let us consider a general string action in $d+1$ dimensions,
\beq
S &=& \frac{1}{4\pi\alpha'} \int_{{\cal M}} d^{2}\xi \sqrt{\hat{g}}
\{G_{MN}(\mbox{\boldmath$X$}) \hat{g}^{ab} \partial_{a}X^{M} 
\partial_{b}X^{N}
+ \alpha' \hat{R} \Phi(\mbox{\boldmath$X$})\} \nonumber \\
& &+ \frac{1}{2\pi\alpha'} \int_{\partial{\cal M}} ds \sqrt{\hat{g}_{ss}}
\alpha' \hat{k} \Phi (\mbox{\boldmath$X$}(s)) \: ,
\eeq
where we have included the boundary term, as it is needed in following
calculation.\footnote
{For the non-linear sigma model with boundaries, see Refs.
\cite{NLS-open} and \cite{H.Osborn}.}
The renormalized action with dimensional regularization up to two loops
 is expressed as
\beq
S &=& \frac{1}{4\pi\alpha'} \int_{{\cal M}} d^{2+\epsilon}\xi \sqrt{\hat{g}}
\left (G_{MN}(\mbox{\boldmath$X$})
+ \a'\frac{1}{\epsilon}C_{MN}(\mbox{\boldmath$X$})\right )
 \hat{g}^{ab} \partial_{a}X^{M} \partial_{b}X^{N} \nonumber \\
& &+
\frac{1}{4\pi\alpha'} \int_{{\cal M}} d^{2+\epsilon}\xi \sqrt{\hat{g}}
 \alpha' \hat{R} \left (\Phi(\mbox{\boldmath$X$})
+\frac{1}{\epsilon}C_{\Phi}(\mbox{\boldmath$X$})\right) \nonumber \\
& &+ \frac{1}{2\pi\alpha'} \int_{\partial{\cal M}} d^{1+\epsilon}s 
\sqrt{\hat{g}_{ss}}
\alpha' \hat{k} 
\left ( \Phi(\mbox{\boldmath$X$}(s))
+\frac{1}{\epsilon}\tilde{C}_{\Phi}(\mbox{\boldmath$X$}(s)) \right),
\eeq
where the symmetric tensor $\frac{1}{\epsilon}C_{MN}$
is the counterterm of the kinetic term, and
the scalar $\frac{1}{\epsilon}C_{\Phi}\;$
($\frac{1}{\epsilon}\tilde{C}_{\Phi}$) 
is the counterterm of
the dilaton term in ${\cal M}\;$ ($\partial{\cal M}$).\footnote
{We implicitly renormalized quadratically divergent terms into the
  cosmological constant term, and we set them to zero.}
If we perform the Weyl transformation
 $\hat{g}_{ab}\mapsto \hat{g}_{ab} e^{\delta\sigma}$
and take the limit of $\epsilon \rightarrow 0$,
 the finite variation of the action at $O ( \delta\sigma)$ is
\beq
\delta_{\sigma} S
&=& \frac{1}{4\pi\alpha'} \int_{{\cal M}} d^{2}\xi \sqrt{\hat{g}}
\alpha' \left \{
\frac{1}{2}\delta\sigma C_{MN}(\mbox{\boldmath$X$})
 \hat{g}^{ab} \partial_{a}X^{M} \partial_{b}X^{N}
+ 
\hat{R} \frac{1}{2}\delta\sigma
C_{\Phi}(\mbox{\boldmath$X$})
\right \} \nonumber \\
& &- 
\frac{1}{4\pi\alpha'} \int_{{\cal M}} d^{2}\xi \sqrt{\hat{g}}
\alpha'  \left( \Phi(\mbox{\boldmath$X$})
               + C_{\Phi}(\mbox{\boldmath$X$}) \right)
 \nabla^{2}(\delta\sigma) \nonumber \\
& &+ 
\frac{1}{2\pi\alpha'} \int_{\partial{\cal M}} ds \sqrt{\hat{g}_{ss}}
\alpha' 
\hat{k} \frac{1}{2}\delta\sigma
\tilde{C}_{\Phi}(\mbox{\boldmath$X$}(s)) \nonumber \\
& &+
\frac{1}{2\pi\alpha'} \int_{\partial{\cal M}} ds \sqrt{\hat{g}_{ss}}
\alpha'
\frac{1}{2}(\hat{n}^{a}\partial_{a}(\delta \sigma))\;
\left( \Phi(\mbox{\boldmath$X$}(s))
+ \tilde{C}_{\Phi}(\mbox{\boldmath$X$}(s))\right),
\label{del-S}
\eeq
where $\hat{n}^{a}$ is the unit outward normal vector.
The term which contains $\nabla^{2}(\delta\sigma)$ in (\ref{del-S})
 is rewritten as
\beq
&-& \frac{1}{4\pi\alpha'} \int_{{\cal M}} d^{2}\xi \sqrt{\hat{g}} \hat{g}^{ab}
\alpha' 
\left \{
 \nabla_{a} \nabla_{b} X^{M}\nabla_{M}
 \left (
  \Phi(\mbox{\boldmath$X$})+C_{\Phi}(\mbox{\boldmath$X$}) \right)
 \right . \nonumber \\
& &  \makebox[5.5cm]{}  +
 \left .
 \partial_{a} X^{M} \partial_{b} X^{N} \nabla_{M}\nabla_{N}
 \Phi(\mbox{\boldmath$X$})
 \right \}
\delta\sigma \nonumber \\
&-&  \frac{1}{4\pi\alpha'} \int_{\partial{\cal M}} ds \sqrt{\hat{g}_{ss}}
\alpha'
(\hat{n}^{a}\partial_{a}(\delta \sigma))\;
\left( \Phi(\mbox{\boldmath$X$}(s))
+ C_{\Phi}(\mbox{\boldmath$X$}(s)) \right ) \makebox[2cm]{}  \nonumber \\
&+&  \frac{1}{4\pi\alpha'} \int_{\partial{\cal M}} ds \sqrt{\hat{g}_{ss}}
\alpha'
\hat{n}^{a}\partial_{a} X^{M} \nabla_{M}
\left( \Phi(\mbox{\boldmath$X$}(s))
+ C_{\Phi}(\mbox{\boldmath$X$}(s)) \right )
\delta\sigma \:,
\label{dilaton_del_sigma}
\eeq
where $M$ and $N$ are the indices of the coordinates of
 spacetime, and $\partial_{N}$ stands for
 $\frac{\partial}{\partial X^{N}}$.

Therefore we get
\beq
\delta_{\sigma} S
&=& \frac{1}{4\pi} \int_{{\cal M}} d^{2}\xi \sqrt{\hat{g}}
\left \{
\left(
   \frac{1}{2} C_{MN}(\mbox{\boldmath$X$})
   -\nabla_{M}\nabla_{N}\Phi(\mbox{\boldmath$X$})
\right)
 \hat{g}^{ab} \partial_{a}X^{M} \partial_{b}X^{N} \right .
 \nonumber \\
& & \makebox[8cm]{}+
\left . 
\hat{R} \frac{1}{2}
C_{\Phi}(\mbox{\boldmath$X$})
\right \}
\delta\sigma  \nonumber \\
& &+
\frac{1}{2\pi} \int_{\partial{\cal M}} ds \sqrt{\hat{g}_{ss}}
\hat{k} \frac{1}{2}
\tilde{C}_{\Phi}(\mbox{\boldmath$X$}(s))
 \delta\sigma  \nonumber \\
& &+ 
\frac{1}{4\pi} \int_{\partial{\cal M}} ds \sqrt{\hat{g}_{ss}}
(\hat{n}^{a}\partial_{a}(\delta \sigma))\;
\left( 
    \tilde{C}_{\Phi}(\mbox{\boldmath$X$}(s))
    -C_{\Phi}(\mbox{\boldmath$X$}(s))
\right) \nonumber \\
& &- \frac{1}{4\pi} \int_{{\cal M}} d^{2}\xi \sqrt{\hat{g}} \hat{g}^{ab} 
 \nabla_{a} \nabla_{b} X^{M}\nabla_{M}
 \left(\Phi(\mbox{\boldmath$X$})+C_{\Phi}(\mbox{\boldmath$X$})\right)
\delta\sigma  \nonumber \\
& &+
\frac{1}{4\pi} \int_{\partial{\cal M}} ds \sqrt{\hat{g}_{ss}}
 \hat{n}^{a}\partial_{a} X^{M} \nabla_{M}
\left( \Phi(\mbox{\boldmath$X$}(s))
+ C_{\Phi}(\mbox{\boldmath$X$}(s)) \right)
\delta\sigma .
\label{dels+gomi}
\eeq
The last three terms cannot be absorbed into any counterterm.
The third term of (\ref{dels+gomi}) goes to zero if
 $\tilde{C}_{\Phi}=C_{\Phi}$. This is
realized if all the $\beta$-functions, except for that of the dilaton,
vanish \cite{H.Osborn}. The problem is to determine how to deal with 
the last two terms.

Fortunately, we can cancel them using field redefinition.
If we perform a field redefinition
 $X^{M} \mapsto X^{M}+\delta X^{M}$,
 the variation of the action is
\beq
\delta_{X} S &=& \frac{1}{4\pi\alpha'}
 \int_{{\cal M}} d^{2}\xi \sqrt{\hat{g}}\;
 2 \hat{g}^{ab}\partial_{a}X^{M} \partial_{b}\delta X_{M} \nonumber \\
& &+
(\mbox{terms proportional to }\delta X_{M}) \nonumber \\
&=& -\frac{1}{4\pi\alpha'}
 \int_{{\cal M}} d^{2}\xi \sqrt{\hat{g}}\;
 2 \hat{g}^{ab}\nabla_{a}\nabla_{b}X^{M} \delta X_{M}  \nonumber \\
& &+
\frac{1}{4\pi\alpha'} \int_{\partial{\cal M}} ds \sqrt{\hat{g}_{ss}}
 2\hat{n}^{a} \partial_{a}X^{M}  \delta X_{M} \nonumber \\
& &+
(\mbox{terms proportional to }\delta X_{M})
.
\label{dxS}
\eeq
Therefore, if we set 
\beq
\delta X^{M}=-\frac{\alpha'}{2}\nabla^{M}
(\Phi(\mbox{\boldmath$X$})+C_{\Phi}(\mbox{\boldmath$X$})) \delta\sigma,
\label{delX}
\eeq
we can cancel the last two terms of (\ref{dels+gomi}).
The remaining terms proportional to $\delta X_{M}$ in (\ref{dxS})
can be absorbed into the counterterms.
Thus, after a proper field redefinition,
$\delta_{\sigma} S +\delta_{X}S$ contains only terms proportional
 to the $\beta$-functions, and we can preserve Weyl invariance
 if we set each of the $\beta$-functions to zero.
We know that the counterterm $C_{\Phi}$ at the one-loop level corresponds 
to the central charge, and is a constant.
Thus the required shift of $X^{M}$ is
\beq
\delta X^{M}=-\frac{\alpha'}{2}\nabla^{M}
\Phi(\mbox{\boldmath$X$}) \delta\sigma.
\label{AdelX-C}
\eeq
Now we emphasize the very important fact that we do {\em not} need field
redefinition at the special point $\mbox{\boldmath$X$}_{0}$ where
$\nabla^{M}\Phi(\mbox{\boldmath$X$}_{0})=0$.

In our model (\ref{actionfinal}), $G_{MN}$, $\Phi$ and the counter terms
depend only on $X^{d+1}=\varphi$, so that (\ref{AdelX-C}) can be 
written as
\beq
\delta \varphi=-\frac{\alpha'}{2}\nabla^{\varphi}
\Phi(\varphi) \delta\sigma 
\eeq
\beq
\delta X^{\mu}=0,
\eeq
where $\mu$ runs from 1 to $d$ and $\nabla^{\varphi}=\nabla^{d+1}$.
Now $G^{\varphi \varphi}=1$, and
 $\nabla^{\varphi}\Phi=\nabla_{\varphi}\Phi$. Thus we obtain
 (\ref{delvarphi}).


\newpage

\begin{thebibliography}{1}


\bibitem{Polchinski:1992vg}
J.~Polchinski,
hep-th/9210045.




\bibitem{Polyakov 1981}
A.M.Polyakov, Phys.Lett.{\bf B103}(1981)207

\bibitem{O.Alvarez}
O.Alvarez, Nucl.Phys.{\bf B216}(1983)125




\bibitem{DK}
J.Distler and H.Kawai, Nucl.Phys.{\bf B321}(1989)509

\bibitem{David}
F.David, Mod.Phys.Lett.{\bf A3}(1988)1651

\bibitem{review of DDK}
H.Kawai, Nucl.Phys.{\bf B}(Proc.Suppl.){\bf 26}(1992)93

\bibitem{Seiberg}
N.Seiberg, Prog.Theor.Phys.Suppl.{\bf 102}(1990)319









\bibitem{boundary state}
M.Li, Phys.Rev.{\bf D54}(1996)1644 (hep-th/9512042);\\
J.Ellis, N.E.Mavromatos and D.V.Nanopoulos,
Int.J.Mod.Phys.{\bf A12}(1997)2639 (hep-th/9605046);\\
P.Mansfield and R.Neves, Nucl.Phys.{\bf B479}(1996)82 (hep-th/9605097);\\
K.Ghoroku,(hep-th/9608020);\\
J.Ambjorn, K.Hayasaka and R.Nakayama,
    J.Mod.Phys.Lett.{\bf A12}(1997)1241 \\(hep-th/9702019);\\
Rui Neves, Phys.Lett.{\bf B411}(1997)73 (hep-th/9706069);\\
A.Rajaraman and M.Rozali, JHEP {\bf 9912} (1999) 0005 (hep-th/9909017);\\
V.Fateev, A.Zamolodchikov and Al.Zamolodchikov, hep-th/0001012

\bibitem{Polchinski 1994}
J.Polchinski, Phys.Rev.{\bf D50}(1994)6041 (hep-th/9407031 )

\bibitem{AG}
E.\'Alvarez and C.G\'omez, Nucl.Phys.{\bf B550}(1999)169 (hep-th/9902012)

\bibitem{NLS-open}
C.Callan, C.lavelace, C.Nappi and S.Yost, Nucl.Phys.{\bf B288}(1987)525;\\
K.Behrndt and D.Dorn, Int.J.Mod.Phys.{\bf A7}(1992)1375;\\
K.Behrndt, Nucl.Phys.{\bf B414}(1994)114 (hep-th/9304096)


\bibitem{H.Osborn}
H.Osborn, Nucl.Phys.{\bf B363}(1991)59



\bibitem{Sen}
A.Sen, hep-th/9904207 and the references therein.

\bibitem{Kawai}
Private communication with H.Kawai.

\bibitem{Green et al}
M.B.Green, Phys.Lett.{\bf B266}(1991)325 and the references therein.
\end{thebibliography}
\end{document}